# Long-range quantum transport of indirect excitons in van der Waals heterostructure

L. H. Fowler-Gerace,[1] Zhiwen Zhou,[1] E. A. Szwed,[1] and L. V. Butov[1]

[1]*Department of Physics, University of California at San Diego, La Jolla, CA 92093, USA*
(Dated: June 3, 2022)

Long lifetimes of spatially indirect excitons (IXs), also known as interlayer excitons, make possible long-range IX propagation. Van der Waals heterostructures composed of atomically thin layers of transition-metal dichalcogenides (TMDs) give an opportunity to realize excitons with high binding energies and provide a materials platform for the realization of both excitonic quantum phenomena and excitonic devices. Propagation of IXs in TMD heterostructures is intensively studied. However, in spite of long IX lifetimes, orders of magnitude longer than lifetimes of spatially direct excitons (DXs), a relatively short-range IX propagation with the $1/e$ decay distances $d_{1/e}$ up to few $\mu$m was reported in the studies of TMD heterostructures. The short-range of IX propagation originates from in-plane potentials, which localize excitons and suppress exciton transport. In particular, significant in-plane moiré potentials predicted in TMD heterostructures can cause an obstacle for IX propagation. In this work, we realize in a MoSe$_2$/WSe$_2$ heterostructure a macroscopically long-range IX propagation with $d_{1/e}$ reaching $\sim 100$ $\mu$m. The strong enhancement of IX propagation is realized using an optical excitation resonant to DXs in the heterostructure. The strong enhancement of IX propagation originates from the suppression of IX localization and scattering and is observed in the quantum regime.

PACS numbers:

A spatially indirect exciton (IX) is a bound pair of an electron and a hole confined in separated layers [1]. Due to the spatial separation of electrons and holes, the lifetimes of IXs can exceed the lifetimes of spatially direct excitons (DXs) by orders of magnitude. The long lifetimes allow IXs to travel long distances before recombination. The long-range IX transport has been extensively studied in GaAs heterostructures where the $1/e$ decay distances of IX luminescence $d_{1/e}$ reach tens and hundreds of microns [2–22]. The long IX propagation distances in GaAs heterostructures allowed uncovering a number of exciton transport phenomena, including the inner ring in exciton emission patterns [5, 7], the exciton localization-delocalization transition in random [5], periodic [10] and moving [14] potentials, the transistor effect for excitons [9], the long-range coherent spin transport [15, 17, 21], and the dissipationless exciton transport over long times evidencing the exciton superfluidity [22].

Excitons exist at temperatures roughly below $E_X/k_B$ ($E_X$ is the exciton binding energy, $k_B$ the Boltzmann constant) [23]. IXs in GaAs heterostructures exist at low temperatures due to their low binding energies: The IX binding energy is typically $\sim 4$ meV in GaAs/AlGaAs heterostructures [24] and $\sim 10$ meV in GaAs/AlAs heterostructures [25]. Furthermore, the temperature of quantum degeneracy, which can be achieved with increasing density before excitons dissociate to electron-hole plasma, scales proportionally to $E_X$ [26]. Therefore, material systems where IXs have high-binding energies can provide the platform for the realization of both high-temperature excitonic quantum phenomena and high-temperature excitonic devices. IXs are explored in III-V and II-VI semiconductor heterostructures based on GaAs [2–22], GaN [27–30], and ZnO [31, 32]. Among these materials, the highest IX binding energy $\sim 30$ meV is in ZnO heterostructures [31].

Excitons with remarkably high binding energies can be realized in van der Waals heterostructures composed of atomically thin layers of transition-metal dichalcogenides (TMDs) [33–36]. The IX binding energies in TMD heterostructures reach hundreds of meV [26, 37], making IXs stable at room temperature [38, 39]. Propagation of both DXs in TMD monolayers [40–46] and IXs in TMD heterostructures [47–55] is intensively studied. However, in spite of long IX lifetimes in TMD heterostructures, orders of magnitude longer than DX lifetimes, a relatively short-range IX propagation with $d_{1/e}$ up to $\sim 3$ $\mu$m was reported in the studies of TMD heterostructures.

The short-range of IX propagation originates from in-plane potentials in the TMD heterostructures. In-plane potentials localize excitons and suppress exciton transport. In addition to a long IX lifetime, a long-range IX propagation requires that the in-plane-potential-induced localization and scattering of IXs is weak. In particular, the long-range IX propagation with a high IX diffusion coefficient is realized in GaAs heterostructures with small in-plane disorder potentials [20].

In contrast to GaAs heterostructures, in mechanically stacked TMD heterostructures, the layers are not perfectly aligned and the misalignment can cause significant moiré superlattice potentials. For MoSe$_2$/WSe$_2$ heterostructures, similar to the heterostructure studied in this work, the IX energy modulations in moiré potentials are predicted to be in the range of tens of meV [56–60]. Moiré superlattices enable studying excitons in in-plane potentials with the period $b \approx a/\sqrt{\delta\theta^2 + \delta^2}$ typically in the $\sim 10$ nm range ($a$ is the lattice constant, $\delta$ the lattice mismatch, $\delta\theta$ the twist angle deviation from $n\pi/3$, $n$ is an integer) [61–66], and the moiré potentials can be affected by atomic reconstruction [67–71]. However, for the exciton propagation, the moiré potentials can cause an obstacle and, along with in-plane disorder potentials, can be responsible for limiting the IX propagation distances to $d_{1/e} \sim 3$ μm in the TMD heterostructures [47–55].

The IX propagation with $d_{1/e} \sim 10$ μm was realized by applying voltage between the top and bottom electrodes in a TMD heterostructure [72]. The origin of this voltage-controlled IX propagation was discussed in terms of the theoretically predicted [59] tuning of the moiré potential by electric-field, enabling the IX delocalization.

In this work, we realize in a MoSe$_2$/WSe$_2$ TMD heterostructure a macroscopically long-range IX propagation with $d_{1/e}$ reaching $\sim 100$ μm. The strong enhancement of IX propagation is realized using an optical excitation resonant to DXs in the heterostructure. The data show that the strong enhancement of IX propagation is caused by the suppression of IX localization and scattering and is observed in the quantum regime.

The MoSe$_2$/WSe$_2$ heterostructure is assembled by stacking mechanically exfoliated 2D crystals on a graphite substrate. The MoSe$_2$ and WSe$_2$ monolayers are encapsulated by dielectric cladding layers of hexagonal boron nitride (hBN). IXs are formed from electrons and holes confined in adjacent MoSe$_2$ and WSe$_2$ monolayers, respectively. No voltage is applied in the heterostructure. In the absence of applied voltage, IXs form the ground state in the MoSe$_2$/WSe$_2$ heterostructure. The heterostructure details are presented in Supporting Information (SI).

The long-range IX propagation with $d_{1/e}$ reaching $\sim 100$ μm (Fig. 1a,b) is realized when the optical excitation has the energy $E_{ex}$ close to the MoSe$_2$ or WSe$_2$ DX energy (Fig. 1e). The IX propagation with even longer $d_{1/e}$ is likely realized for $E_{ex}$ close to the DX energy, however, the finite heterostructure dimensions limit the longest $d_{1/e}$, which can be reliably established, to $\sim 100$ μm (Fig. 1a,b). In contrast, for a non-resonant excitation, the range of IX propagation is substantially shorter (Fig. 1a,b). The IX luminescence is traced along the entire IX propagation path from the excitation spot to the edge of the heterostructure (Fig. 1c). Spatial intensity modulations along the IX propagation will be considered elsewhere.

The factors, which contribute to the enhancement of IX propagation, are outlined below. IXs have built-in electric dipoles $\sim ed_z$ ($d_z$ is the separation between the electron and hole layers) and the interaction between IXs is repulsive. An enhancement of IX transport with increasing IX density due to the repulsive IX interaction is caused (1) by the suppression of IX localization and scattering and (2) by the IX-interaction-induced drift from the origin.

The data show that the first factor causes the major effect. The nature of the second factor is an increase of IX energy at the excitation spot $\delta E$ with increasing IX density $n$ (Fig. 1f) that causes IX drift from the origin [7]. The nearly resonant excitation produces a higher $n$ due to a higher absorption, thus increasing $\delta E$ and, in turn, the IX drift. However, the higher $n$ and $\delta E$ can be also achieved for nonresonant excitation using higher excitation powers $P_{ex}$. Figure 1g shows $d_{1/e}$ vs. $\delta E$ both for the resonant and nonresonant excitation. For the same $\delta E$, a much higher $d_{1/e}$ is realized for the resonant excitation. This shows that the effect of IX energy increase at the excitation spot on the enhancement of IX propagation is minor. The strong enhancement of IX propagation at resonant excitation originates from the suppression of localization and scattering of IXs.

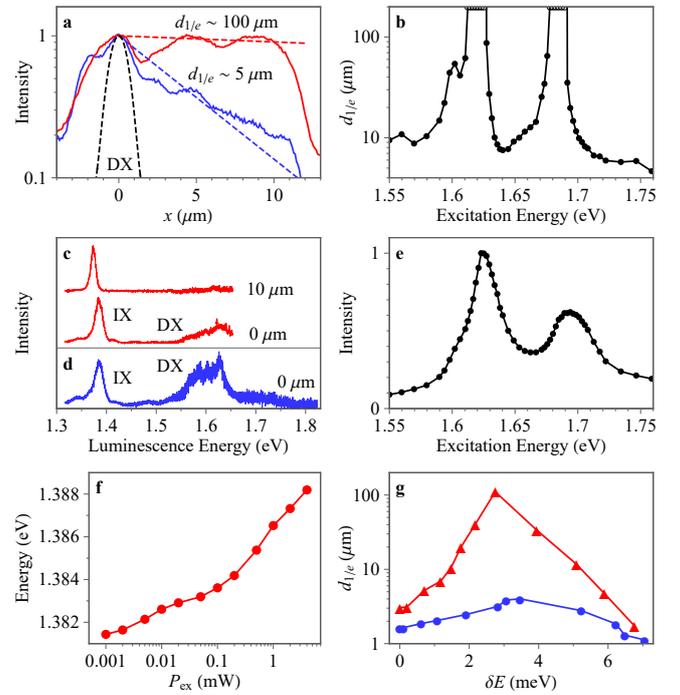

FIG. 1: The long-range IX propagation realized using optical excitation resonant to DXs. (a) Normalized IX luminescence profiles for laser excitation off ($E_{ex}$ = 1.771 eV, blue) and near ($E_{ex}$ = 1.676 eV, red) the DX absorption resonance. The heterostructure active area extends from $x = -3$ to 12 μm. The blue and red dashed lines show decays with $d_{1/e}$ = 5 and 100 μm, respectively. The black dashed line shows the DX luminescence profile in the WSe$_2$ monolayer, this profile is close to the laser excitation profile for a short DX propagation. (b) The $1/e$ decay distance of IX luminescence $d_{1/e}$ vs. $E_{ex}$. $d_{1/e}$ are calculated from fitting the IX luminescence profiles to exponential decays in the region $x = 0 - 11$ μm. The data with the fit indicating $d_{1/e} > 100$ μm are presented by points on the edge. The finite heterostructure dimensions limit the longest $d_{1/e}$, which can be reliably established, to $\sim 100$ μm. (c,d) The spectra for resonant [$E_{ex}$ = 1.694 eV, (c)] and non-resonant [$E_{ex}$ = 1.96 eV, (d)] excitation. The relative intensity of the higher-energy DX luminescence is lower for the resonant excitation, evidencing a lower temperature of the exciton system. (e) Integrated IX luminescence intensity vs. $E_{ex}$ showing two absorption peaks corresponding to the MoSe$_2$ and WSe$_2$ DXs. For data in (a-e), $P_{ex}$ = 0.2 mW, $T$ = 1.7 K. (f) The IX energy at the excitation spot vs. $P_{ex}$. $E_{ex}$ = 1.623 eV, $T$ = 6 K. (g) $d_{1/e}$ vs. the IX energy increase at the excitation spot $\delta E$ for resonant ($E_{ex}$ = 1.623 eV, red) and nonresonant ($E_{ex}$ = 1.96 eV, blue) excitation. $T$ = 6 K. For all data, the $\sim 1.5$ μm laser spot is centered at $x = 0$.

The IX energy increase due to the repulsive interaction $\delta E$ (Fig. 1f) can be used for estimating the IX density $n$. For instance, for the optimal IX propagation conditions, that is for the nearly resonant excitation (Fig. 1) and the temperature and $P_{ex}$ corresponding to the long $d_{1/e}$ (Fig. 3), $\delta E \sim 3$ meV (Fig. 1f) and an estimate for $n$ using the mean-field "plate capacitor" formula $\delta E = nu_0$ [73] gives $n \sim 2 \times 10^{11}$ cm$^{-2}$ ($u_0 = 4\pi e^2 d_z/\varepsilon$, $d_z \sim 0.6$ nm, the dielectric constant $\varepsilon \sim 7.4$ [74]). This estimate can be improved by taking into account IX correlations that increases $n$ in comparison to the mean-field estimate. The increase is $\sim 3$ times for IXs in GaAs heterostructures [10, 75]. While similar estimates for correlations in TMD heterostructures are yet unavailable, using the correlation correction similar to that in GaAs gives the estimated IX density $n \sim 6 \times 10^{11}$ cm$^{-2}$ for the optimal IX propagation conditions. In turn, an estimated temperature of quantum degeneracy $T_d = (2\pi\hbar^2 n)/(k_B m) \sim 30$ K [26] ($m \sim 0.9 m_0$ is the IX mass in the TMD heterostructure [76, 77]). This estimate shows that the observed strong enhancement of IX propagation at resonant excitation due to the suppression of IX localization and scattering occurs (i) in the quantum regime and (ii) in a dilute IX gas with the densities below the Mott transition density $n_{Mott} \gtrsim 10^{12}$ cm$^{-2}$ [26, 78].

In the quantum regime, the pioneering works on IXs predict IX superfluidity [1] that suppresses the IX localization and scattering. However, a theory of quantum transport of IXs in TMD heterostructures, in particular in the presence of moiré potentials, yet need to be developed. Therefore, below, we use the classical drift-diffusion model [7], outlined in SI, to discuss the characteristics of IX propagation.

For instance, within the drift-diffusion model of IX transport [7], the suppression of IX localization and scattering is accounted for by screening of the in-plane potential by repulsively interacting IXs, see SI. The screening of in-plane potential is more effective when the excitation is close to the DX resonances. In particular, heating of the exciton system by the nearly resonant excitation is, in general, smaller than for non-resonant excitation, and the colder IXs can screen the in-plane potential more effectively [7, 10]. The most efficient screening of the in-plane potential by IXs is realized below the temperature of IX condensation [79].

Figure 2 shows the kinetics of IX propagation from the excitation spot for the nearly optimal propagation conditions. The IX kinetics is measured during the rectangular-shaped laser excitation pulses with the du-ration 100 ns and period 300 ns. The 200 ns off time exceeds the IX lifetime (Fig. 4) and is sufficient for a substantial decay of the IX signal. The IX luminescence at locations away from the excitation spot is delayed in comparison to the IX luminescence in the excitation spot (Fig. 2a,b). The delay times $\delta t$ for the IX cloud to expand to the locations separated by distance $R$ from the origin allow estimating the IX transport characteristics. Fitting $R$ vs. $\delta t$ by $R^2 \sim D^* \delta t$ (Fig. 2d), gives an estimate for the effective IX diffusion coefficient $D^* \sim 70$ cm$^2$/s.

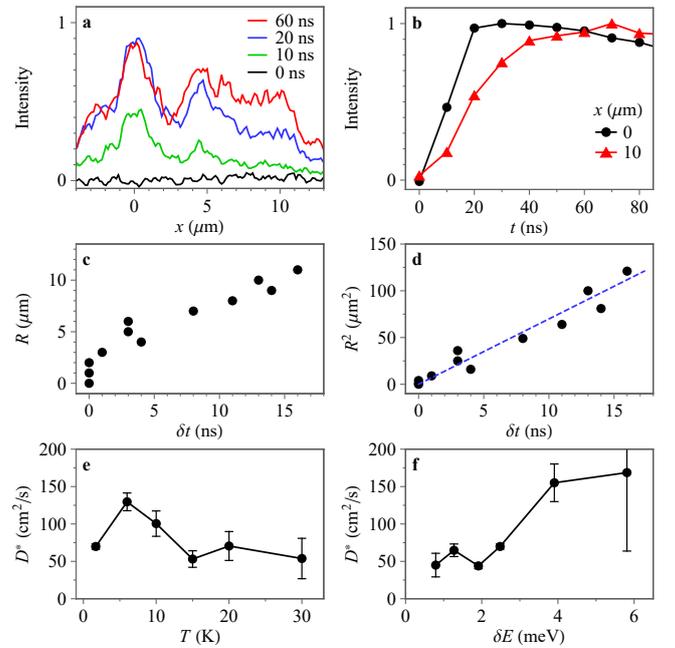

FIG. 2: Time-resolved IX propagation. (a) IX luminescence profiles during the laser excitation pulse at different times. (b) Normalized IX luminescence intensity vs. time at different positions. (c) $R$ vs. $\delta t$. $\delta t$ is the time to reach 70% of the maximum intensity at distance $R$ from the excitation spot relative to that time in the excitation spot. (d) $R^2$ vs. $\delta t$. The dashed line is a fit to the data with $D^* = R^2/\delta t = 70$ cm$^2$/s. (e,f) $D^*$ vs. temperature (e) and vs. $\delta E$ (f). $P_{ex} = 0.15$ mW for data in (a-e). $T = 1.7$ K for data in (a-d,f). For all data, the times are given at the ends of the 10 ns signal integration windows, the excitation pulse starts at $t = 0$, $E_{ex} = 1.694$ eV, the $\sim 2$ $\mu$m laser spot is centered at $x = 0$.

For a diffusive IX propagation, both IX drift and diffusion contribute to the expansion of IX cloud. With increasing distance from the origin, both the IX density $n$ and IX interaction energy $\delta E \sim nu_0$ decrease. The IX energy gradient causes the IX drift away from the origin. Fitting an IX cloud expansion by $R^2 \sim D^* \delta t$ probes the effective IX diffusion coefficient $D^* = D + \mu n u_0$, which includes both the diffusion and drift due to the density gradient [80]. The IX mobility $\mu$ can be estimated using the Einstein relation $\mu = D/k_B T$, giving $D^* = D(1 + nu_0/k_B T)$. For $nu_0 \sim 3$ meV at $P_{ex} = 0.15$ mW (Fig. 1f), $D^* \sim 70$ cm$^2$/s (Fig. 2d), and $T = 1.7$ K, this equation gives an estimate for the IX diffusion coefficient $D \sim 4$ cm$^2$/s. In turn, the estimated IX mobility $\mu = D/k_B T \sim 3 \times 10^4$ cm$^2$/(eV s).

The data $R$ vs. $\delta t$ (Fig. 2c) indicate that an IX cloud expansion occurs with a nearly constant velocity after first few ns. The estimated average velocity of the IX cloud expansion for the time range $\Delta t = 1 - 20$ ns, $v = \Delta R/\Delta t \sim 5 \times 10^4$ cm/s (Fig. 2c).

The variation of $D^*$ with temperature and $\delta E$ is presented in Fig. 2e,f ($\delta E$ characterizes the IX density $n$ at the origin). An enhancement of $D^*$ with density can be in principle explained by the classical IX drift and diffusion: both the exciton diffusion coefficient $D$ and the IX drift should increase with $n$, the former due to the enhanced screening of in-plane potential and the latter due to the enhanced $\delta E$ at the origin [7], see SI. An enhancement of IX propagation with density due to these factors was observed in earlier studies of both GaAs and TMD heterostructures, see e.g. Refs. [7, 10, 48, 54, 55]. As outlined above, the long-range IX propagation at resonant excitation, presented here, originates from the suppression of IX localization and scattering in the quantum dilute IX gas and is qualitatively different from the IX propagation in earlier studies of TMD heterostructures.

A reduction of $D^*$ with temperature is observed for $T \gtrsim 6$ K (Fig. 2e). A suppression of IX propagation with temperature is consistent with the expected behavior for quantum exciton transport [1] and complies with the reduction of superfluid density with temperature: A nearly linear reduction of the superfluid density with temperature at temperatures below the Berezinskii-Kosterlitz-Thouless transition is found in the theory [81]. We note however that more accurate measurements are needed for a quantitative comparison with the theory. This forms a subject for future works. In particular, an increase of $D^*$ with temperature at $T \lesssim 6$ K (Fig. 2e) need to be understood. In general, the contributions of both the normal and superfluid components to the quantum exciton transport should be explored. The classical drift-diffusion model [7] is inconsistent with the observed temperature dependence of IX transport, see SI.

The excitation power and temperature dependence of IX propagation is presented in Fig. 3. With increasing $P_{ex}$, the IX propagation distance $d_{1/e}$ changes nonmonotonically. A relatively short-range IX propagation is observed at the lowest $P_{ex}$ (Fig. 3a,d). This is consistent with the IX localization in the in-plane potential. Increasing $P_{ex}$ and, in turn, IX density $n$ suppresses the IX localization and scattering and results in an enhancement of IX propagation. The longest IX propagation is achieved at $P_{ex} \sim 0.2$ mW and the further enhancement of $P_{ex}$ leads to the suppression of IX propagation (Fig. 3b,d). This suppression is related to the reduction of IX lifetime at high $P_{ex}$ (Fig. 4c).

With increasing temperature, the IX propagation first enhances, reaches maximum around $T \sim 6$ K, and reduces at higher temperatures (Fig. 3e). This behavior is related to the temperature dependence of $D^*$ (Fig. 2e), with the suppression at high temperatures further enhanced by the reduction of IX lifetime (Fig. 4d). The measured $P_{ex} - T$ and derived $n - T$ diagrams for the IX propagation distance $d_{1/e}$ are shown in Fig. 3f and SI, respectively.

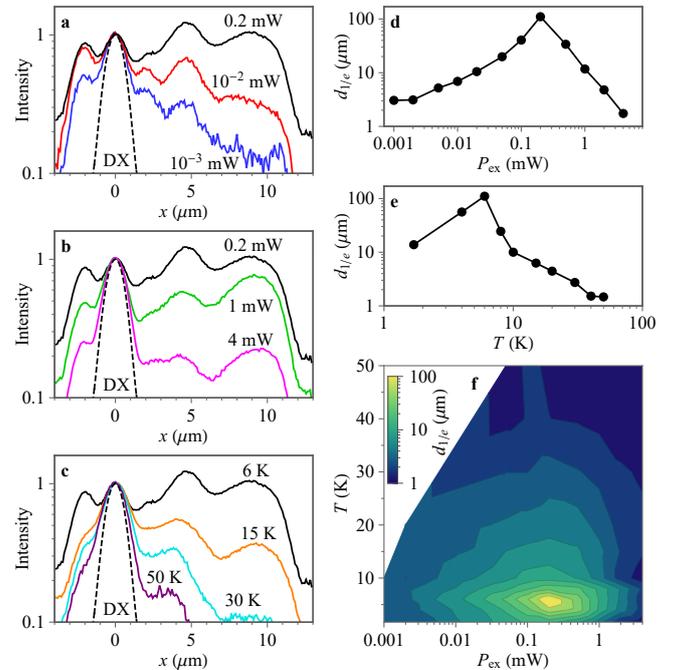

FIG. 3: Excitation power and temperature dependence of IX propagation. (a-c) Normalized IX luminescence profiles for different $P_{ex}$ (a,b) and temperatures (c). The dashed line shows the DX luminescence profile in the WSe$_2$ monolayer, this profile is close to the laser excitation profile for a short DX propagation. $T = 6$ K (a,b,d), $P_{ex} = 0.2$ mW (c,e). (d,e) The $1/e$ decay distance of IX luminescence $d_{1/e}$ vs. $P_{ex}$ (d) and vs. temperature (e). (f) Contour plot showing the decay distance $d_{1/e}$ vs. both $P_{ex}$ and temperature. The greatest IX propagation occurs at $T \sim 6$ K and $P_{ex} \sim 0.2$ mW. For all data, $E_{ex} = 1.623$ eV, the $\sim 1.5$ $\mu$m laser spot is centered at $x = 0$.

Figure 4 shows the IX luminescence decay kinetics after the laser excitation pulse is off. The IX decay times $\tau$ (Fig. 4) are orders of magnitude longer than the DX decay times [82]. Both increasing $P_{ex}$ (Fig. 4a,c) and temperature (Fig. 4b,d) lead to a reduction of $\tau$. This reduction of $\tau$ contributes to suppressing the IX propagation at high excitation powers and temperatures as outlined above.

In summary, a macroscopically long-range IX propagation with the decay distances $d_{1/e}$ reaching $\sim 100$ $\mu$m is observed in a van der Waals MoSe$_2$/WSe$_2$ heterostructure. The strong enhancement of IX propagation is realized using an optical excitation resonant to DXs in the heterostructure. The strong enhancement of IX propagation originates from the suppression of IX localization and scattering and is observed in the quantum regime.

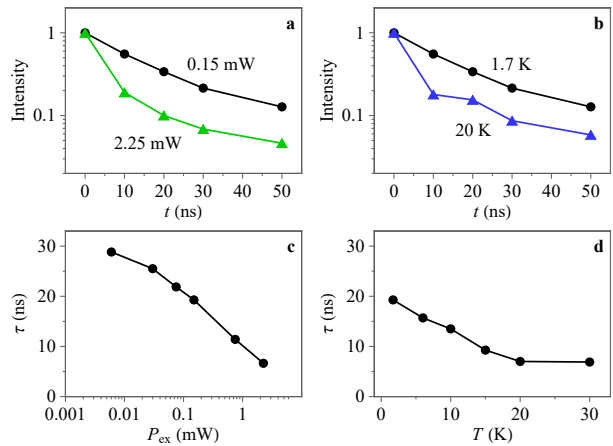

FIG. 4: Excitation power and temperature dependence of IX decay kinetics after the excitation pulse end. (a,b) Spatially integrated IX luminescence intensity vs. time for different $P_{ex}$ (a) and temperatures (b). The times are given at the ends of the 10 ns signal integration windows. The excitation pulse ends at $t = 0$. (c,d) The IX lifetime $\tau$ vs. $P_{ex}$ (c) and vs. temperature (d). $\tau$ is the initial decay time after the excitation pulse end. $T = 1.7$ K (a,c), $P_{ex} = 0.15$ mW (b,d). For all data, $E_{ex} = 1.694$ eV.

**ACKNOWLEDGMENTS**

We thank Darius Choksy and Michael Fogler for discussions. These studies were supported by DOE Office of Basic Energy Sciences under Award No. DE-FG02-07ER46449. The heterostructure fabrication and data analysis were supported by NSF Grant No. 1905478.

# Supporting Information for
# Long-range quantum transport of indirect excitons in van der Waals heterostructure

L. H. Fowler-Gerace,[1] Zhiwen Zhou,[1] E. A. Szwed,[1] and L. V. Butov[1]

[1]*Department of Physics, University of California at San Diego, La Jolla, CA 92093, USA*
(Dated: June 3, 2022)

PACS numbers:

### The heterostructure details

The van der Waals heterostructure was assembled using the dry-transfer peel-and-lift technique [1]. Crystals of hBN, MoSe$_2$, and WSe$_2$ were first mechanically exfoliated onto different Si substrates that were coated with a double polymer layer consisting of polymethyl glutarimide (PMGI) and polymethyl methacrylate (PMMA). The bottom PMGI was then dissolved with the tetramethylammonium hydroxide based solvent CD-26, causing the top PMMA membrane with the target 2D crystal to float on top of the solvent. The PMMA membrane functions both as a support substrate for transferring the crystal and as a barrier to protect the crystal from the solvent. Separately, a large graphite crystal was exfoliated onto an oxidized Si wafer, which served as the basis for the heterostructure. The PMMA membrane supporting the target crystal was then flipped over and aligned above a flat region of the graphite crystal using a micromechanical transfer stage. The two crystals were brought into contact and the temperature of the stage was ramped to 80° C in order to increase adhesion between the 2D crystals. Then, the PMMA membrane was peeled off leaving the bilayer stack on the wafer. The procedure was repeated leading to a multicrystal stack with the desired layer sequence. Sample annealing was performed by immersing the sample in Remover PG, an N-methyl-2-pyrrolidone (NMP) based solvent stripper, at 70° C for 12 hours.

No intentional sample doping was done; however, unintentional $n$–type doping is typical for TMD layers [1]. The thickness of bottom and top hBN layers is about 40 and 30 nm, respectively. The MoSe$_2$ layer is on top of the WSe$_2$ layer. The long WSe$_2$ and MoSe$_2$ edges reach ∼ 30 and ∼ 20 $\mu$m, respectively, which enables a rotational alignment between the WSe$_2$ and MoSe$_2$ monolayers. The twist angle estimated from the angle between the long WSe$_2$ and MoSe$_2$ edges $\delta\theta = 0.5° \pm 0.8°$ (Fig. S1).

Figure S1 presents a microscope image showing the layer pattern of the heterostructure. The layer boundaries are indicated. The hBN layers cover the entire areas of MoSe$_2$ and WSe$_2$ layers. There was a narrow graphene electrode on the top of the heterostructure around $x = 2$ $\mu$m for $y = 0$, Fig. S1. This electrode was detached. The IX luminescence reduction around $x = 2$ $\mu$m can be related with residual graphene layers on the heterostructure.

The discussions in this work apply to both R and H (AA and AB) stacking; therefore, we did not verify stacking in the sample. Our work shows that the long-range quantum transport of IXs with suppressed IX localization and scattering can be realized in TMD heterostructures using an optical excitation with the energy close to the DX energy. The studies of sample statistics, and in particular, verifying the role of the rotational alignment between crystals on the IX propagation, is the subject for future works.

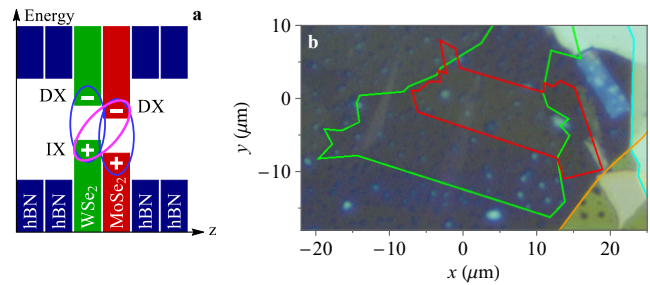

FIG. S1: (a) Schematic energy-band diagram for the heterostructure. The ovals indicate a direct exciton (DX) and an indirect exciton (IX) composed of an electron (–) and a hole (+). (b) A microscope image showing the layer pattern of the heterostructure. The green, red, cyan, and orange lines indicate the boundaries of WSe$_2$ and MoSe$_2$ monolayers and bottom and top hBN layers, respectively.

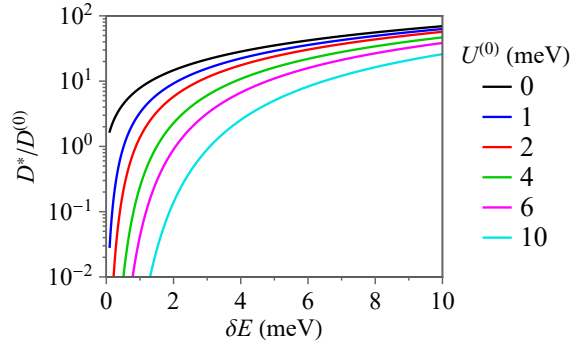

FIG. S2: $D^*/D^{(0)}$ vs. $\delta E$ for different $U^{(0)}$. $T = 1.7$ K.

**Optical measurements**

In the cw experiments, excitons were generated by a cw Ti:Sapphire laser with tunable excitation energy or a cw HeNe laser with excitation energy $E_{ex}$ = 1.96 eV. Luminescence spectra were measured using a spectrometer with resolution 0.2 meV and a liquid-nitrogen-cooled CCD.

The IX luminescence kinetics was measured using a pulsed semiconductor laser with $E_{ex}$ = 1.694 eV nearly resonant to WSe$_2$ DX energy. The emitted light was detected by a liquid-nitrogen-cooled CCD coupled to a PicoStar HR TauTec time-gated intensifier.

The experiments were performed in a variable-temperature 4He cryostat. The sample was mounted on an Attocube xyz piezo translation stage allowing adjusting the sample position relative to a focusing lens inside the cryostat.

**The drift-diffusion model of IX transport**

The drift-diffusion model of IX transport [2–4], used for the estimates for IX diffusion coefficient and mobility in the main text, is outlined in this section. Within this model, IX transport is described by the equation for IX density $n$

$$\frac{\partial n}{\partial t} = \nabla \left[ D \nabla n + \mu n \nabla (u_0 n) \right] + \Lambda - \frac{n}{\tau} \quad (1)$$

The first and second terms in square brackets in Eq. 1 describe IX diffusion and drift currents, respectively. The latter originates from the IX repulsive dipolar interactions and is approximated by the mean-field "plate capacitor" formula for the IX energy shift with density $\delta E = n u_0$, $u_0 = 4\pi e^2 d_z / \varepsilon$ [5]. The diffusion coefficient

$$D = D^{(0)} \exp[-U^{(0)}/(k_B T + n u_0)] \quad (2)$$

accounts for the temperature- and density-dependent screening of the long-range-correlated in-plane potential landscape by interacting IXs, $D^{(0)}$ is the diffusion coefficient in the absence of in-plane potential and $U^{(0)}/2$ is the amplitude of the in-plane potential [2–4]. The IX mobility $\mu$ is given by the Einstein relation $\mu = D/(k_B T)$. The IX generation rate $\Lambda$ has a profile of the laser excitation spot. $\tau$ is the IX lifetime.

Both the IX-interaction-induced screening of in-plane potential and the IX-interaction-induced drift from the origin contribute to the enhancement of IX transport with increasing IX density $n$. In particular, within the classical drift-diffusion model [2–4], the enhancement of IX transport due to the IX-interaction-induced screening of in-plane potential is described by Eq. 2, while the enhancement of IX transport due to the IX-interaction-induced drift from the origin is described by Eq. 3 as outlined in the main text

$$D^* = D[1 + n u_0/(k_B T)]. \quad (3)$$

Equations 2 and 3 show that within the classical drift-diffusion model [2–4], both $D$ and $D^*$ should increase with density for any $U^{(0)}$ (Fig. S2). The variation of $D^*$ increases with $U^{(0)}$ (Fig. S2), and the variation of $D^*$ in the experiment (~ 3 times for $\delta E = 1 - 4$ meV, Fig. 2f) corresponds to vanishing in-plane potential ($U^{(0)} < 1$ meV). The predicted $U^{(0)}$ for MoSe$_2$/WSe$_2$ heterobilayers with R and H stacking is ~ 100 and ~ 25 meV, respectively [6, 9].

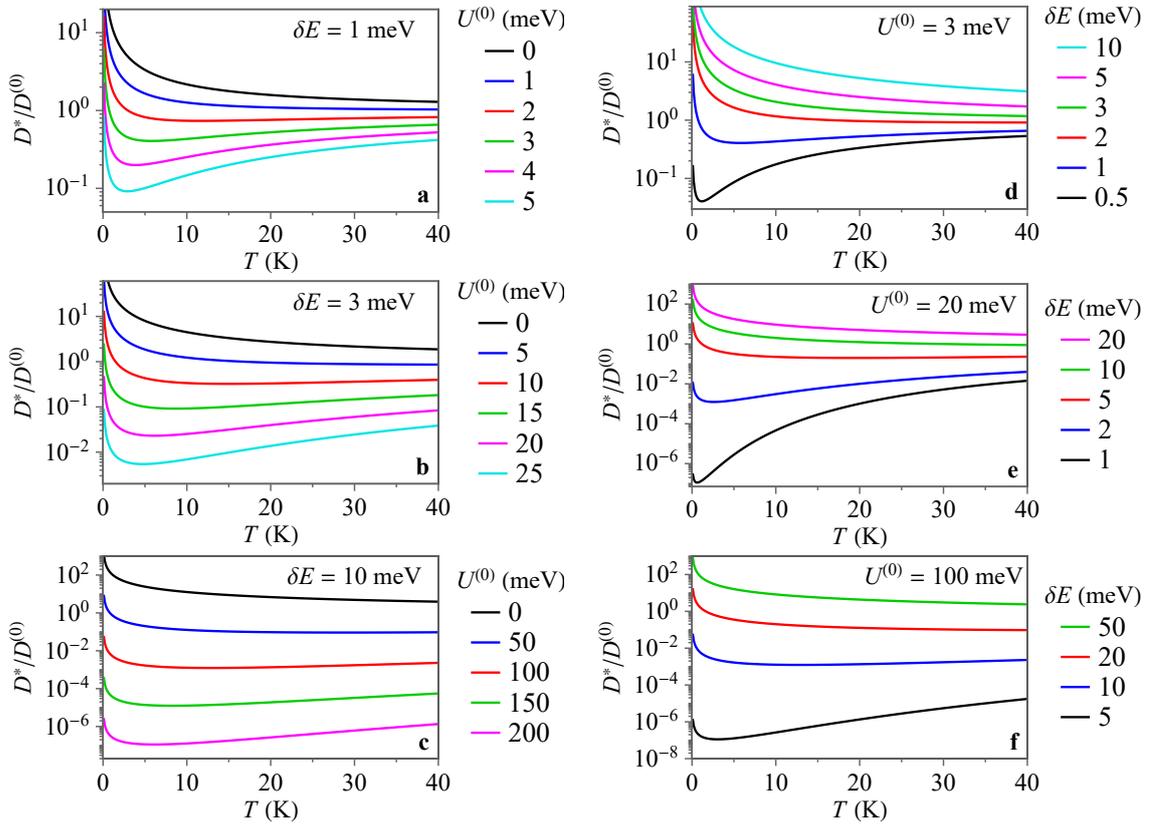

FIG. S3: $D^*/D^{(0)}$ vs. $T$ for different $\delta E$ and $U^{(0)}$.

Equations 2 and 3 show that within the classical drift-diffusion model [2–4], $D$ should increase with temperature for any $U^{(0)}$, however, the dependence of $D^*$ on temperature is nonmonotonic (Fig. S3). Within this model, $D^*$ increases with temperature when $U^{(0)} > \delta E(1 + \delta E/k_B T)$. The classical drift-diffusion model [2–4] (Eqs. 1-3, Fig. S3) is inconsistent with the observed temperature dependence of IX transport (Fig. 2e): If $D^*$ increases with temperature at $T \sim 2 - 6$ K, as in the experiment, then within the model $D^*$ should also increase with temperature at higher temperatures [if $U^{(0)} > \delta E(1 + \delta E/k_B T)$ at lower $T$, this inequality holds for higher $T$, see e.g. the black lines in Fig. S3d,e,f] that is inconsistent with the experiment; If $D^*$ reduces with temperature at $T \gtrsim 6$ K, as in the experiment, then within the model $D^*$ should also reduce with temperature at lower temperatures [if $U^{(0)} < \delta E(1 + \delta E/k_B T)$ at higher $T$, this inequality holds for lower $T$, see e.g. the black lines in Fig. S3a,b,c)] that is inconsistent with the experiment.

The drift-diffusion Eq. 1 can be supplemented by the thermalization equation, which describes heating of excitons by photoexcitation and cooling via interaction with phonons, and by including the heterostructure in-plane potential in the drift term [3, 4]. In particular, the exciton thermalization can lead to the appearance of the inner ring in exciton luminescence patterns [3, 4]. The Einstein relation can be extended to the generalized Einstein relation $\mu = D(e^{T_d/T} - 1)/(k_B T_d)$, which gives $\mu = D/(k_B T)$ for the classical drift and diffusion [2–4].

**IX interaction energies and the long-range IX propagation**

Below, we briefly discuss the IX interaction energies $\delta E$ corresponding to the long-range IX propagation. In GaAs heterostructures, the strong enhancement of IX propagation, the IX delocalization, is observed when the IX interaction energy becomes comparable to the amplitude of the in-plane potential, which is, in turn, comparable to the IX luminescence linewidth at low IX densities [11]. For the long-range IX propagation in the MoSe$_2$/WSe$_2$ heterostructure (Fig. 3), the IX interaction energy $\delta E \sim 3$ meV (Fig. 1f). This value is comparable to the smallest IX linewidth $\sim 4$ meV at low IX densities in the MoSe$_2$/WSe$_2$ heterostructure (Fig. S4), in qualitative agreement with the data in GaAs heterostructures [11].

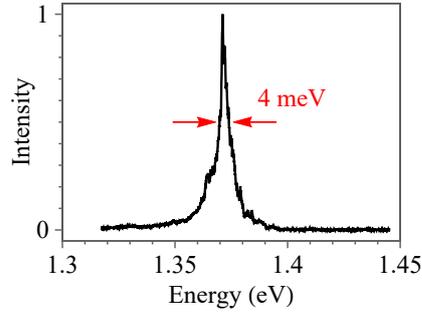

FIG. S4: IX luminescence spectrum at a low IX density in the MoSe$_2$/WSe$_2$ heterostructure. $P_{ex}$ = 0.005 mW, $T$ = 1.7 K, $E_{ex}$ = 1.96 eV.

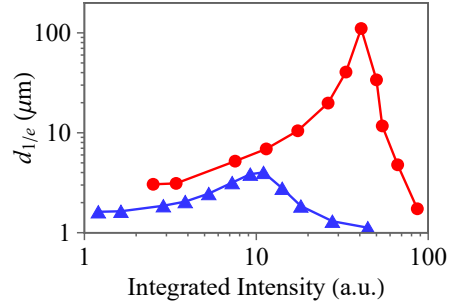

FIG. S5: $d_{1/e}$ vs. integrated IX intensity in the entire MoSe$_2$/WSe$_2$ heterostructure. Excitation is either nonresonant ($E_{ex}$ = 1.96 eV, blue), or near resonant ($E_{ex}$ = 1.623 eV, red) to the DX. $T$ = 6 K, the ∼ 1.5 $\mu$m laser spot is centered at $x$ = 0.

However, this value, $\delta E \sim 3$ meV, is significantly smaller than the predicted IX energy modulations in moiré superlattice potentials in MoSe$_2$/WSe$_2$ heterostructures that are in the range of tens of meV [6–10]. Theoretical understanding of the long-range IX quantum transport in the moiré potentials forms a subject for future work.

**IX propagation decay distance for nearly resonant excitation and nonresonant excitation**

The nearly resonant excitation produces a higher IX density and, in turn, a stronger IX luminescence signal due to a higher absorption (Fig. 1). The higher IX densities can be also achieved by nonresonant excitation with higher excitation powers $P_{ex}$. Figure 1g shows that for the same $\delta E$, a much higher $d_{1/e}$ is realized for the resonant excitation and that the strong enhancement of IX propagation at resonant excitation originates from the suppression of localization and scattering of IXs. This is further confirmed by Fig. S5, which also compares the IX propagation decay distances $d_{1/e}$ for the nearly resonant and nonresonant excitations. Figure S5 shows that in the regime of the long-range IX propagation, for the same IX signal in the heterostructure, achieved by a higher $P_{ex}$ for nonresonant excitation, $d_{1/e}$ is significantly longer in the case of the nearly resonant excitation. This indicates that the nearly resonant excitation is essential for the achievement of the long-range IX propagation.

**Rough estimates of the radiative and nonradiative lifetime variation with $P_{ex}$ and temperature**

The IX radiative and nonradiative lifetimes, $\tau_r$ and $\tau_{nr}$, can be roughly estimated from the measured IX total luminescence intensity $I$ and decay time $\tau$. Figure S6 shows $\tau_r$ and $\tau_{nr}$ estimated using $\tau_r^{-1} = (I/\Lambda)\tau^{-1}$ and $\tau_{nr}^{-1} = \tau^{-1} - \tau_r^{-1}$. Within this estimate, the IX luminescence kinetics is approximated as monoexponential. It is also assumed that there is no IX spatial escape from the system. This is approximated by integrating the IX luminescence signal over the heterostructure. For an estimate of the generation rate $\Lambda$, we use $I_D = \alpha \Lambda$, where $I_D$ is the DX luminescence signal for the excitation at the MoSe$_2$ monolayer outside the heterostructure and $\alpha$ is the quantum efficiency for this DX luminescence. Qualitatively similar variations of $\tau_r$ and $\tau_{nr}$ with $P_{ex}$ and temperature are obtained for different values of $\alpha$ (Fig. S6). These estimates suggest that the reduction of $\tau$ at the high excitation powers and temperatures is mainly caused by the enhanced nonradiative recombination (Fig. S6).

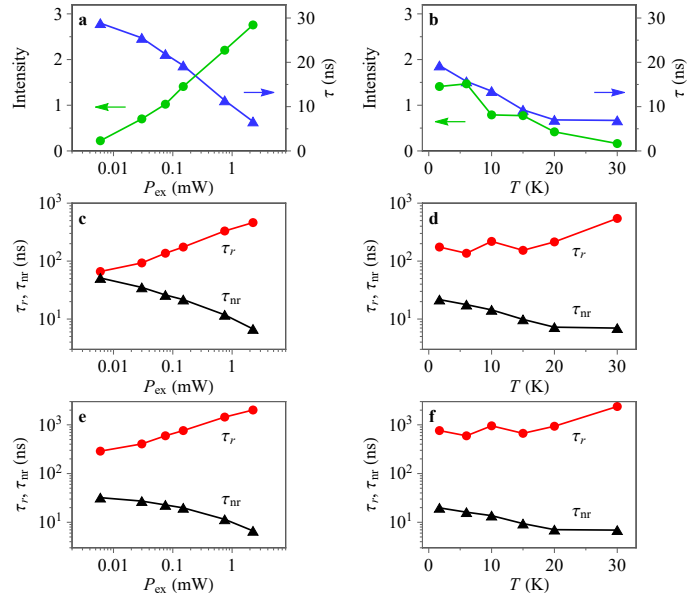

FIG. S6: Rough estimate of radiative and nonradiative lifetimes, $\tau_r$ and $\tau_{nr}$. (a,b) Spatially integrated IX luminescence intensity $I$ (green) and IX lifetime $\tau$ (blue) vs. $P_{ex}$ (a) and temperature (b). $T = 1.7$ K (a), $P_{ex} = 0.15$ mW (b), and $E_{ex} = 1.694$ eV (a,b). $\tau$ is the initial decay time after the excitation pulse end. (c-f) Estimated $\tau_r$ and $\tau_{nr}$ vs. $P_{ex}$ (c,e) and temperature (d,f) for the quantum efficiency of DX luminescence in the MoSe$_2$ monolayer $\alpha = 100\%$ (c,d) and 25% (e,f).

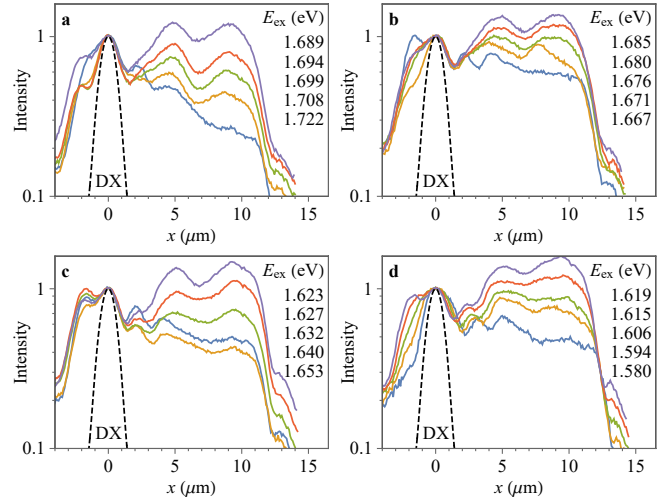

FIG. S7: Normalized IX luminescence profiles for different laser excitation energies $E_{ex}$. The dashed line shows the DX luminescence profile in the WSe$_2$ monolayer, this profile is close to the laser excitation profile for a short DX propagation. $P_{ex} = 0.2$ mW, $T = 1.7$ K, the ~ 1.5 $\mu$m laser spot is centered at $x = 0$.

**IX luminescence profiles for different laser excitation energies**

Normalized IX luminescence profiles for different laser excitation energies $E_{ex}$ are shown in Fig. S7. The long-range IX propagation with high decay distances $d_{1/e}$ is realized for $E_{ex}$ close to the MoSe$_2$ or WSe$_2$ DX energy (Fig. 1). For some of the IX luminescence profiles the IX luminescence intensity increases with separation from the origin and no decay is observed within the finite heterostructure dimensions. The description of this intensity increase is beyond Eq. 1. It may be obtained by supplementing Eq. 1 by the thermalization equation, which describes heating of excitons by photoexcitation and cooling via interaction with phonons, and including the heterostructure in-plane potential in the drift term [3, 4]. In particular, the exciton thermalization can lead to the appearance of the inner ring in exciton luminescence patterns, that is an exciton luminescence intensity increase with separation from the origin [3, 4]. The data with the fit indicating $d_{1/e} > 100$ μm and, in particular, showing the IX luminescence intensity increase with separation from the origin, are presented in Fig. 1b by points on the edge.

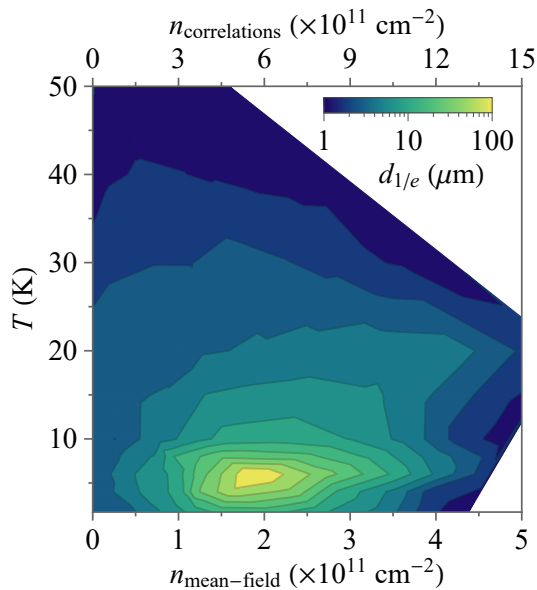

FIG. S8: Contour plot showing the decay distance $d_{1/e}$ vs. both density and temperature. $E_{ex}$ = 1.623 eV.

### $n - T$ diagram for the IX propagation distance $d_{1/e}$

$n - T$ diagram for the IX propagation distance $d_{1/e}$ is shown in Fig. S8. This diagram is obtained from the $P_{ex} - T$ diagram (Fig. 3f). The IX density is estimated from the IX energy shift (Fig. 1f) using the mean-field "plate capacitor" formula $\delta E = n u_0$ ($n_{\text{mean-field}}$ in Fig. S8) and with the correlation correction similar to that in GaAs heterostructures as outlined in the main text ($n_{\text{correlations}}$ in Fig. S8).